\documentclass[aps, prb, preprint, %
               showpacs, groupedaddress, amsmath, amssymb]{revtex4}
\usepackage{graphicx}
\usepackage{bm}

\newcommand{\fig}[1]{FIG.~\ref{#1}}
\renewcommand{\vec}[1]{\mathbf{#1}}

\begin{document}

\title{Thermal conduction of carbon nanotubes using molecular dynamics}

\author{Zhenhua Yao}
\email{smayzh@nus.edu.sg}
\author{Jian-Sheng Wang}
\altaffiliation{also in Department of Computational Science, National
  University of Singapore, Singapore 117543}
\affiliation{Singapore-MIT Alliance, National University of Singapore,
  Singapore 117576}

\author{Baowen Li}
\affiliation{Department of Physics, National University of Singapore,
  Singapore 119260}

\author{Gui-Rong Liu}
\affiliation{
  Department of Mechanical Engineering, National University of
  Singapore, Singapore 119260}


\date{\today}

\begin{abstract}
  The heat flux autocorrelation functions of carbon nanotubes (CNTs)
  with different radius and lengths is calculated using equilibrium
  molecular dynamics. The thermal conductance of CNTs is also
  calculated using the Green-Kubo relation from the linear response
  theory. By pointing out the ambiguity in the cross section
  definition of single wall CNTs, we use the thermal conductance
  instead of conductivity in calculations and discussions. We find
  that the thermal conductance of CNTs diverges with the CNT length.
  After the analysis of vibrational density of states, it can be
  concluded that more low frequency vibration modes exist in longer
  CNTs, and they effectively contribute to the divergence of thermal
  conductance.
\end{abstract}

\pacs{61.48.+c, 63.22.+m, 66.70.+f, 68.70.+w}

\keywords{thermal conduction, carbon nanotube, molecular dynamics,
  Green-Kubo}

\maketitle

\section{Introduction}
\label{sec:intro}

The carbon nanotube (CNT) was discovered by S.~Iijima in 1991
\cite{Iijima}. Since then, its unique mechanical, electrical and
optical properties initiated intensive research on this quasi
one-dimensional material \cite{cnt review}. CNTs have high Young
modulus and strength \cite{yao2001}, as well as high thermal
conductivity. Many novel applications in various areas have been
proposed, including nanoscale electronic devices in the next
generation electronic technologies.

As the dimensions of electronic devices shrink to nanoscale, the
thermal conduction problem becomes more and more important, as a
significant energy may be dissipated in a compact space. However, it
is very difficult to measure the thermal conducting ability of
nanoscale devices.  Furthermore, Fourier Law, which describes the
macroscopic thermal conduction phenomena, may not be appropriate for
low dimensional systems. Therefore, it is important to study the
thermal conduction of nanoscale systems and to develop theoretical and
computational methods for predicting the thermal properties of
nanoscale materials and devices.

There are mainly two approaches to theoretically study the thermal
conduction phenomena of nanoscale materials, the first is a
macroscopic method using continuum models and kinetic theories, such
as Boltzmann transport equation \cite{boltzmann,kubo}, the second is a
fundamental microscopic method based on first principle atomistic
simulations or quantum mechanical models. In the second approach,
various methods are proposed to model the physical system and
calculate the thermal conductivity. These methods include equilibrium
and non-equilibrium molecular dynamics (MD) simulation as well as
model-coupling theory, etc \cite{lepri}.  These methods model the
physical system from scratch and make no empirical assumption.

The understanding of heat conduction and development of a complete
theory is a long-standing and formidably difficult task. For
insulating crystals, the problem of heat transportation by lattice
vibrations is still far from be solved from some points of view.  For
mathematical simplicity, one-dimensional or two-dimensional lattices
of atoms are naturally considered \cite{lepri}. This issue has been
addressed for several decades. Recently, Baowen Li \textit{et al}
established a connection between anomalous heat conduction and
anomalous diffusion in one-dimensional systems \cite{libw}, and
Jian-Sheng Wang \textit{et al} studied the anomalous thermal
conduction in 1D chains by using MD and mode-coupling
theory \cite{wangjs}.

In CNTs generally two physical mechanisms contribute to the thermal
conduction:
\begin{enumerate}
\item Electron--phonon interactions, which mainly depend on electronic
  band structures of and the electron scattering process, etc.
\item Phonon--phonon interactions, which depend on the vibrational
  modes of the lattice.
\end{enumerate}

For semiconductor CNTs in room temperature, phonon--phonon interactions
dominate the overall thermal conductivity, and electron--phonon
interactions have only a small contribution due to the large band gap
and low density of free charge carriers \cite{kelly}.  Fortunately
this part contribution to thermal conduction can be well studied by
using classical MD.

The phonon mean free path in the axial direction of CNTs is estimated
to be very long, about 100 nm --- 1$\mu$m, and much longer than that
of other materials as well as the size of simulation domain, thus the
thermal conductivity of CNTs which are shorter than a few $\mu$m may
have ballistic transport features \cite{maruyama}. On the other hand,
the finite size effect constrains the phonon motion and causes the
thermal conductance seems variable with the CNT length. Actually, it
is difficult to make the simulation domain larger than phonon mean
free path even on supercomputers, thus finding the ``correct'' value
of thermal conductance remains a difficult task.

In the last few years, there are many research activities on this
subject. Berber \textit{et al} studied the thermal conductivity
$\kappa$ of CNTs and the dependence of $\kappa$ on temperature and
suggested that $\kappa$ is about 6600 W/mK for CNT (10,10) at room
temperature \cite{berber}. J.~Che \textit{et al} calculated the
thermal conductivity of diamond materials and CNTs, and showed that
the theoretical value of thermal conductivity converges as the
simulation system size increases. However, in their papers the errors
of thermal conductivity values are too large to draw an accurate
conclusion \cite{che}.  S.~Maruyama \textit{et al} studied the heat
conduction in finite length CNTs using non-equilibrium MD and
calculated the thermal conductivity from the measured temperature
gradients and energy budgets in phantom molecules, and claimed that
thermal conductivity of CNT $(5,5)$ diverges as the power law where
the power index is 0.32 \cite{maruyama}.  M.~A.~Osman \textit{et al}
calculated the temperature dependence of the thermal conductivity and
found that $\kappa$ shows a peaking behavior before falling off at
higher temperatures due to the onset of Umklapp scattering
\cite{osman}.  S.~G.~Volz investigated the thermal conductivity of
bulk silicon crystals based on MD simulation using Stillinger--Weber
potential, and found that $\kappa$ is independent to the length $L_x$
of nanowire when $L_x$ is larger than 16 lattice constants and the
cross section area is smaller than a certain value \cite{volz}.

In addition to these theoretical research there are some experimental
work on the heat conduction of CNTs. D.~Yang \textit{et al}
investigated the thermal conductivity of multi-wall CNTs using a
pulsed photo-thermal reflectance technique, and suggested that the
effective $\kappa$ could be about 200 W/mK \cite{djyang}. P.~Kim and
L.~Shi \textit{et al} measured the thermal conductivity of single CNT
using a microfabricated suspended device and found that $\kappa >
3000$ W/mK at room temperature \cite{lishi}. J.~Hone \textit{et al}
measured the temperature-dependent thermal conductivity of crystalline
ropes of single-wall CNTs and argued that $\kappa$ is dominated
by phonons at all temperatures \cite{hone}.

In this work we use the Green--Kubo relation derived from linear
response theory to examine the thermal conductance by calculating the
heat flux autocorrelation functions. However, finite size artifacts
are still involved due to the frequency cutoff and the artificial
autocorrelation introduced by periodic boundary conditions, which
is consistent with the results of S.~Volz \cite{volz}. We find that the
lowest frequency of lattice vibration modes is limited by the size
of simulation domain, and the thermal conductance of an infinite long
CNT maybe infinite.

\section{Computation of thermal conductivity using MD}
\label{sec:comp}

\subsection{Green-Kubo relation and heat flux}
\label{sec:green-kubo}

In macroscopic model of thermal conduction, the thermal conductivity
is defined from Fourier's law which is for heat flow under non-uniform
temperature distribution. And the heat flux $\vec{j}$ can be defined
as $\vec{j} = -\kappa \nabla T$, where $\kappa$ is the thermal
conductivity tensor and $T$ is the temperature distribution.

From the intuition of Fourier's law, a simple approach to study the
thermal conduction of CNTs is putting the two ends in two heat
reservoirs with different temperature (usually $T_0+\Delta T$ and
$T_0-\Delta T$, where $T_0$ is supposed to be the average temperature
of the system), and measuring the heat flux along the axial direction
and then calculating the thermal conductance. In simulations the heat
flux should be collected after the system becomes steady, and a large
number of averages over time are needed to get smooth temperature
gradient curve and accurate heat flux data. However, the simulation
domain which MD can efficiently handle is not large enough, and the
temperature gradient due to reasonable temperature difference of two
heat reservoirs (please note that too small temperature difference
gives large error and poor results) is far too large to be realistic.
Moreover, as the thermal conductance strongly depends on the
temperature, calculation results from non-uniform temperature
distribution may not be accurate. P.~Schelling \textit{et al}
systematically compared the equilibrium and non-equilibrium methods for
computing the thermal conductivity of insulating materials
\cite{schelling} and mentioned these problems.

Due to aforementioned reasons, in this work we use the
fluctuation--dissipation theorem from linear response theory which
connects the energy dissipation to the thermal fluctuations in
equilibrium state \cite{kubo, hoover}. In this method, the thermal
conductivity in axial direction of CNTs can be expressed in terms
of heat flux autocorrelation function \cite{kubo, hoover},
\begin{equation}
  \label{eq:kappa}
  \kappa = \frac{1}{k_{\text{B}}T^2 V}\int_0^{\infty}
  \langle J(t)J(0) \rangle dt,
\end{equation}
where $J(t) = \int \vec{j}(\vec{r},t)dV$ is the total heat flux
in axial direction, and $V$ is the volume of the system. From the
local energy balance equation
\begin{equation}
  \label{eq:ene-balance}
  \frac{\partial \epsilon(\vec{r},t)}{\partial t} +
  \nabla \cdot \vec{j}(\vec{r},t) = 0,
\end{equation}
where $\epsilon(\vec{r},t)$ is the energy density (i.e., energy per
unit volume), and note that $\epsilon(\vec{r},t)$ and
$\vec{j}(\vec{r},t)$ are macroscopic concepts, a microscopic
expression for total heat flux can be derived as follows,
\begin{equation}
  \label{eq:heatflux}
  \vec{J}(t) = \frac{d}{dt}\sum_i \vec{r}_i(t)\epsilon_i(t),
\end{equation}
where $\vec{r}_i(t)$ is the time-dependent coordinate of atom $i$. In
MD simulation, the total potential energy can be divided among atoms,
then the site energy $\epsilon_i(t)$ can be taken to be
\begin{equation}
  \label{eq:eps(t)}
  \epsilon_i = \frac{1}{2}m_i\vec{v}^2 + \sum_j u(r_{ij}).
\end{equation}

In the above equation, $u(r_{ij})$ is in fact a many-body potential
\cite{tersoff}, and the calculation of total heat flux $\vec{J}(t)$ is
much more complicated in this case than in case of using a simple
pairwise potential function.

\subsection{Interatomic potential}
\label{sec:potential func}

Currently there are several choices of potential functions for
describing interatomic interactions in carbon materials: Tersoff
potential which was published in 1989 for the latest parameters
\cite{tersoff}; Brenner potential which was originally published in
1990 \cite{brenner1} and revised in 2002 \cite{brenner2};
Environment-Dependent Interaction Potential for carbon materials by
N.~A.~Marks published in 2000 \cite{marks}; and a new bond order
potential which parameters are fitted to tight-binding results
\cite{bond-order}. In these potentials, the latter two haven't been
widely recognized, Brenner potential with latest parameters gives
accurate results and is widely used. However, it is observed that in
long-time micro-ensemble running\footnote{We run over $10^8$ steps, as
  a large number of steps is needed for obtaining accurate heat flux
  autocorrelation function data in our work.}, Brenner potential gives
larger total energy deviation than Tersoff potential does due to its
complicated interpolation functions.  On the other hand, Tersoff
potential is stable in long-time running according to our tests and
gives fairly accurate results. Q.~Zheng \textit{et al} compared
Tersoff and Brenner potentials in their theoretical analysis to the
thermal conduction of single-wall CNTs \cite{qzheng} and got good
results using both potentials, Berber has also used Tersoff potential
to study the thermal properties of CNTs and calculated the thermal
conductivity \cite{berber}. Therefore, in our simulation we use
Tersoff potential.

In the past decades, Tersoff potential has been widely used to study
the mechanical and thermodynamic properties of silicon and carbon
materials, and results obtained are accurate and can be comparable to
those from first principle methods, such as from tight-binding method
and density functional theory.  Tersoff potential can be formally
written as a summation of pairwise interactions,
\begin{equation}
  \label{eq:uij}
  V_{\text{tot}} = \frac{1}{2} \sum_{ij} \left\{
  f_{\text{c}}(r_{ij})\left[ V_{\text{R}}(r_{ij}) - 
    B_{ij}V_{\text{A}}(r_{ij}) \right] \right\},
\end{equation}
where $V_{\text{R}}$ and $V_{\text{A}}$ are the repulsive and
attractive parts of the potential, and their functional forms are
given below,
\begin{equation}
  \label{eq:VRA}
  V_{\text{R}}(r) = A \exp(-\lambda r), \qquad
  V_{\text{A}}(r) = B \exp(-\mu r),
\end{equation}
\begin{equation}
  \label{eq:fc}
  f_{\text{c}}(r) = \begin{cases}
    1 & r<R\\
    \frac{1}{2}\left[ 1+\cos\frac{\pi(r-R)}{S-R} \right] & R \le r \le S \\
    0 & r>S
    \end{cases},
\end{equation}
where $f_{\text{c}}(r)$ is a cutoff function which explicitly
restricts the interactions within the nearest neighbors, and
dramatically reduces the redundant computation in the force/potential
evaluation procedure.  In Eq.~\eqref{eq:uij} $B_{ij}$ is a bond order
parameter and depends on the bonding environment around atom $i$ and
$j$. $B_{ij}$ implicitly contains multi-body information and thus
the whole potential function is actually a multiple body potential. The
function form of $B_{ij}$ can be written as follows,
\begin{equation}
  \label{eq:bij}
  B_{ij} = \left[ 1+(\beta \zeta_{ij})^n \right] ^{-\frac{1}{2n}}.
\end{equation}
The detailed information and parameters of
Eq.~\eqref{eq:uij}$\sim$\eqref{eq:bij} are given in Tersoff's paper
\cite{tersoff}.

\subsection{Finite size effect}
\label{sec:finite size}

One of major concerns in simulation of CNTs to calculate the thermal
conductivity is the finite size effect due to periodic boundary
condition applied in the axial direction. As the simulation is
conducted in a periodic box, the long wavelength vibration mode of
lattice is cut off while the CNT is short. This effect causes that a
short CNT's thermal conductivity is smaller than a long CNT's. By
using MD, we investigate the thermal conductivity of CNTs with
different length as well as the relationship between the thermal
conductivity and the length, to study its convergence with system
size.

According to above discussions, a large system has more long
wavelength vibration modes, and correspondingly has longer phonon mean
free path. Thus for longer CNTs, the calculated thermal conductivity
will be larger due of contribution of long wavelength vibration modes.

\subsection{Thermal conductance vs thermal conductivity}
\label{sec:thermal conductance}

It should be mentioned here that since an isolated single-wall CNT's
cross section can be defined in different ways, its thermal
conductivity has also arbitrary definitions and is not a well defined
quantity. Some definitions of the cross section of single-wall CNT
are:
\begin{enumerate}
\item Consider CNT as a solid cylinder, then the cross section area will be
  $\pi R^2$, where $R$ is the radius of CNT;
\item Consider CNT as a hollow cylinder, then the cross section area
  will be $2 \pi R \delta$, where $\delta$ is the thickness of CNT
  shell. In literature usually two values of $\delta$ are used, one is
  $3.4$ \AA, which is the inter-layer distance of graphite materials,
  the other is $1.42$ \AA, which is the length of $sp^2$ bond.
\end{enumerate}

Therefore, in literature many different values of thermal conductivity
are reported, some of them mainly differ in the cross section
definition.

Obviously, the definition of the cross section are is not important
for the thermal conduction research of CNTs, as we only need to
calculate and compare the results consistently. However, for comparing
different results from different research groups, this arbitrary in
the thermal conductivity calculation must be eliminated. In this work,
we use the quantity of ``thermal conductance'', which equals to
thermal conductivity times cross section area. Thus, the thermal
conduction has the dimension of ``Watt$\cdot$meter/Kelvin''.

\subsection{Simulation procedure}
\label{sec:simu details}

In this work, CNTs with different size are investigated.  Firstly,
armchaired CNT $(10, 10)$ with different length are simulated, then
CNT $(15, 15)$ and $(5, 5)$ are simulated. In our MD simulation
program, time integration algorithm is implemented by using velocity
Verlet method. For improving the computation performance, a new
neighbor list algorithm using cell decomposition is employed
\cite{yao}. In all simulation cases, periodic boundary condition is
used only in the axial direction of CNTs. For each simulation case, we
carry out the following three steps:
\begin{enumerate}
\item Firstly canonical ensemble MD is running for $10^5 \sim 5\times
  10^5$ steps in order to take the average system temperature to 300 K
  and wait until system reaches thermal equilibrium.
\item Then followed by micro-canonical ensemble running for another
  $10^5 \sim 5\times 10^5$ steps and wait until system reaches a thermal
  equilibrium in new ensemble.
\item Finally micro-canonical ensemble MD continues to run and heat
  flux data are calculated and collected in every time step. After every
  $10^5$ steps, the power spectrum of heat flux data are online
  calculated, meanwhile, its arithmetic average and Fourier transform,
  which is heat flux autocorrelation function, as well as the
  statistical errors are calculated and dumped to disk files.
\end{enumerate}

In this work, the last step runs infinitely and stops until accurate
results are obtained after many times of average.  Generally $10^8$
steps were carried out in this step, in other words, about $1000$
averages have been done to obtain the final data. Total amount of CPU
time is about three months on 10 Pentium III 866MHz PCs and three
dual-CPU Alpha EV67/667MHz workstations.

In MD simulation, time step is $0.8$ fs, and the canonical ensemble
simulation is implemented by using No\'se-Hoover algorithm
\cite{nose}.

\section{Results and discussions}
\label{sec:results}

\begin{figure}[htbp]
  \centering
  \includegraphics[width=0.4\textwidth]{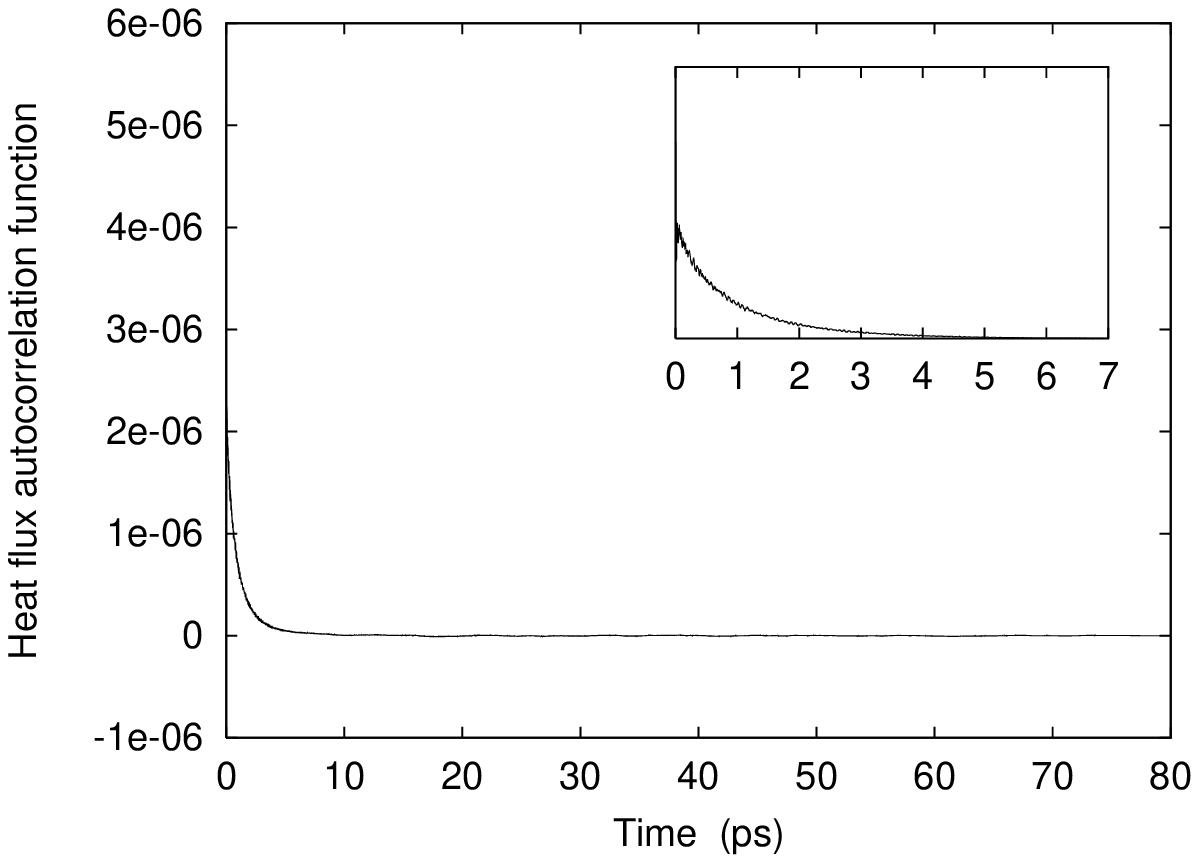}
  \includegraphics[width=0.4\textwidth]{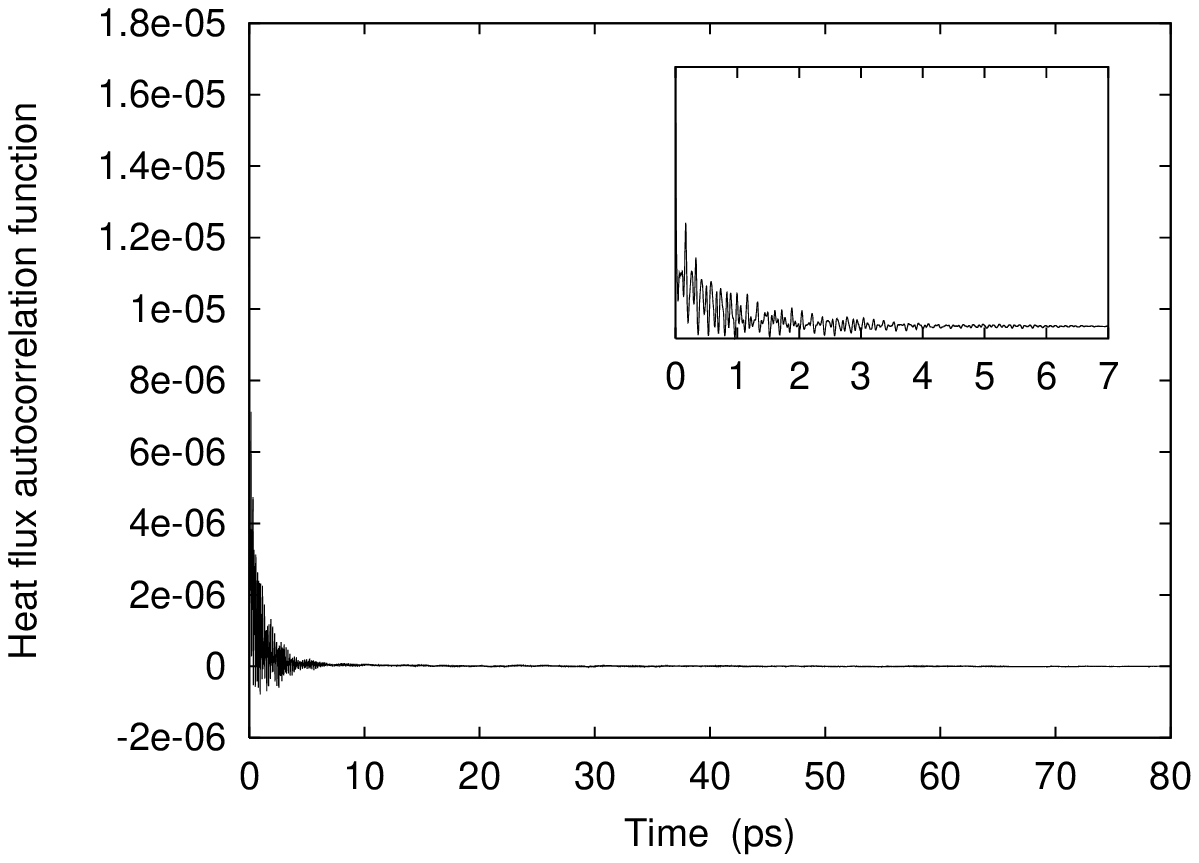}
  \includegraphics[width=0.4\textwidth]{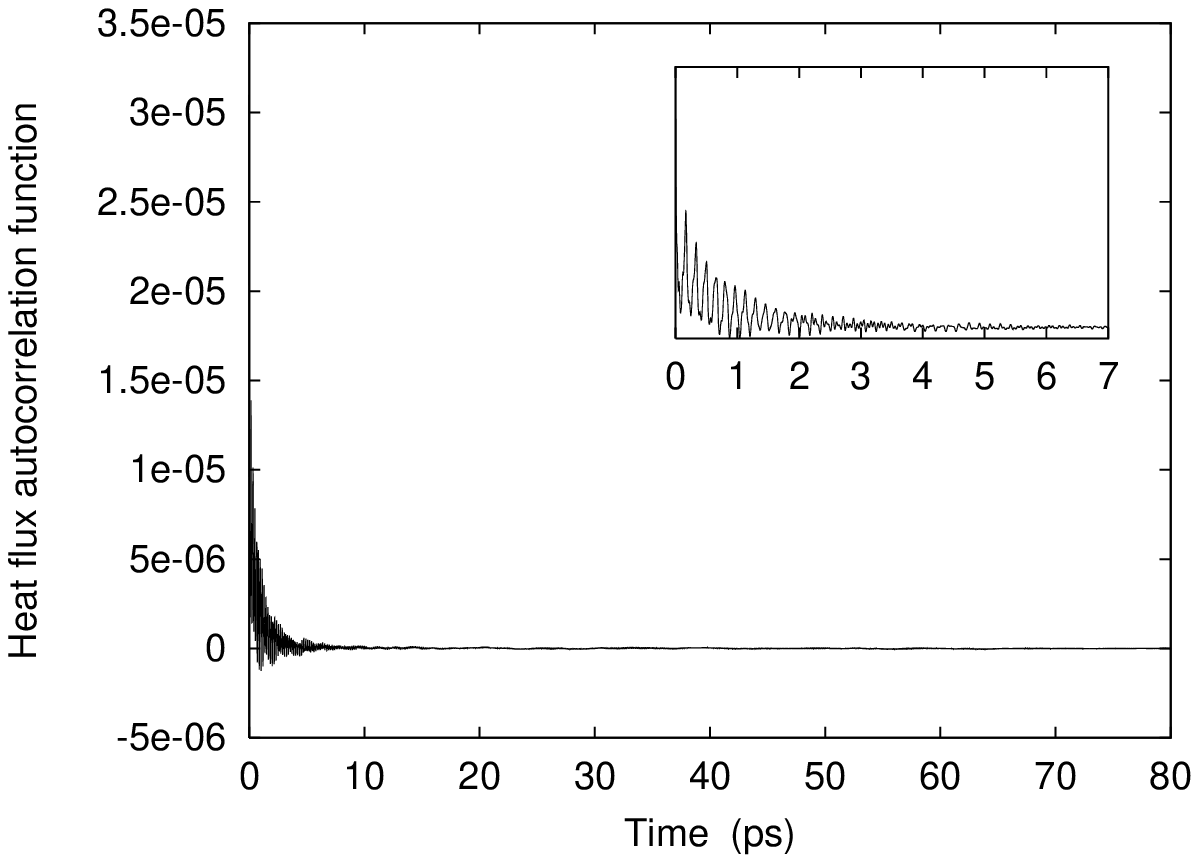}
  \includegraphics[width=0.4\textwidth]{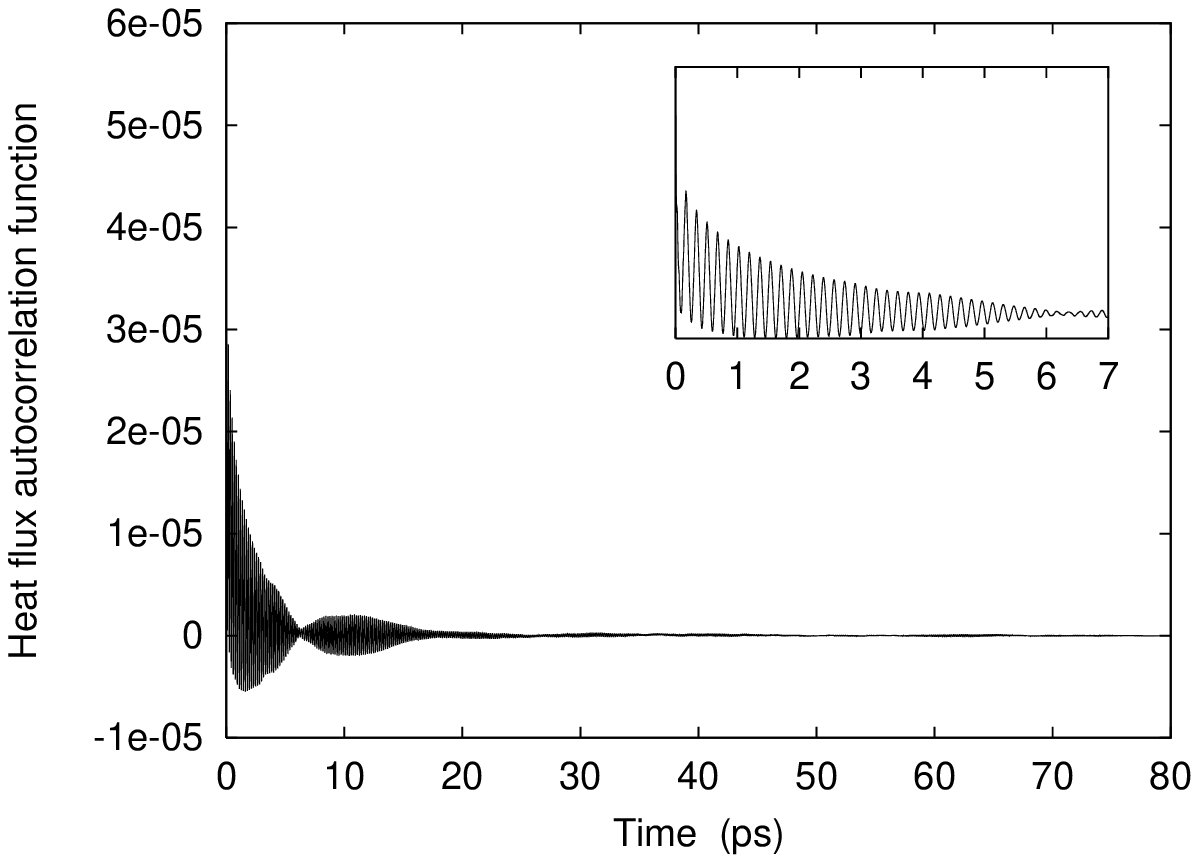}
  \caption{Heat flux autocorrelation function of CNT $(10,10)$ 
    with 50, 100, 200 and 400 layers, respectively (from left to
    right, top to bottom). Insets are initial decay in initial 8000
    steps.}
  \label{fig:hfacf}
\end{figure}

\fig{fig:hfacf} shows the initial autocorrelation function of the
total heat flux along the axial direction of CNT $(10,10)$ with 50,
100, 200 and 400 layers, and a very sharp decay in the beginning and
following a very slow decay can be seen clearly. An oscillation in
autocorrelation function can also be seen in the curves and it becomes
larger when the CNT is longer. It can also been seen an increase
of autocorrelation function values as the CNT length increases. The
fast initial decay is believed to be contributed by high frequency
vibrational modes in the CNT, and slow decay is contributed by low
frequency modes which have much longer wavelength.

\begin{figure}[htbp]
  \centering
  \includegraphics[width=0.4\textwidth]{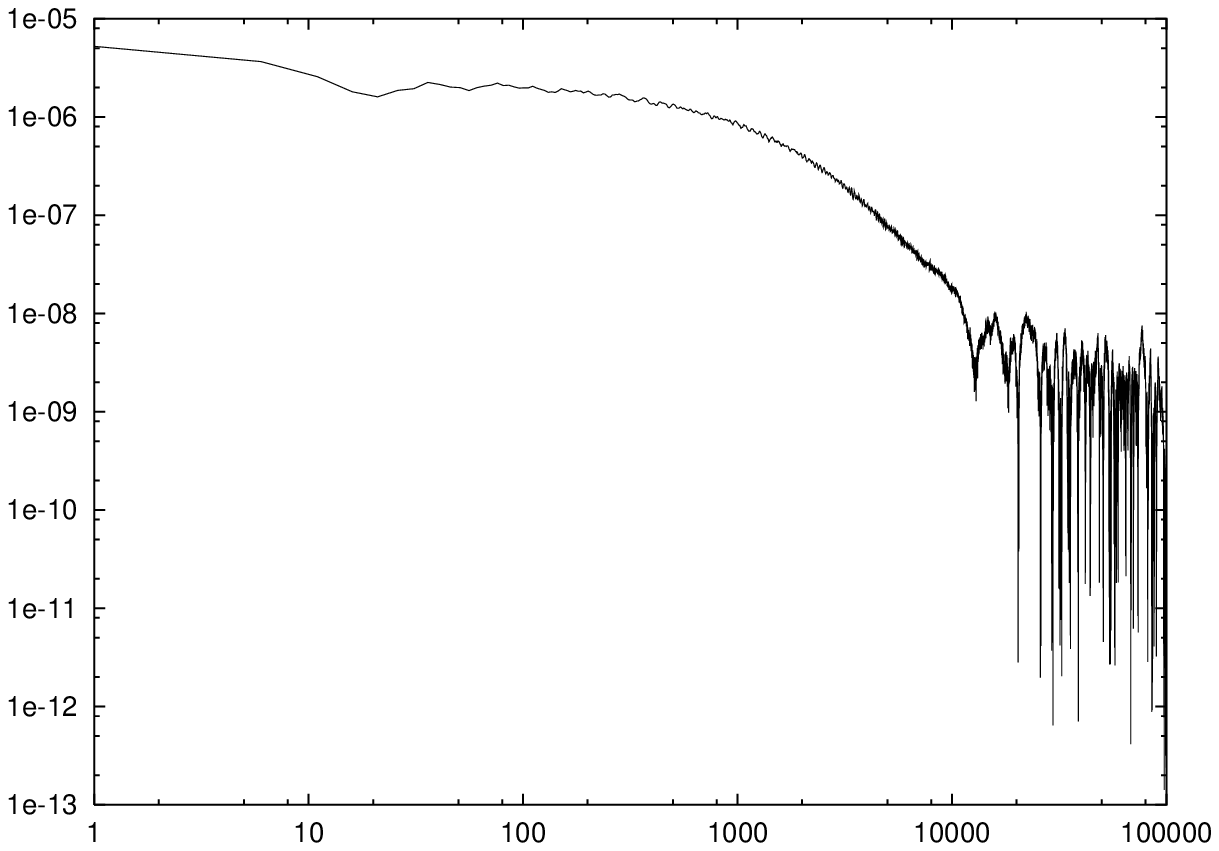}
  \includegraphics[width=0.4\textwidth]{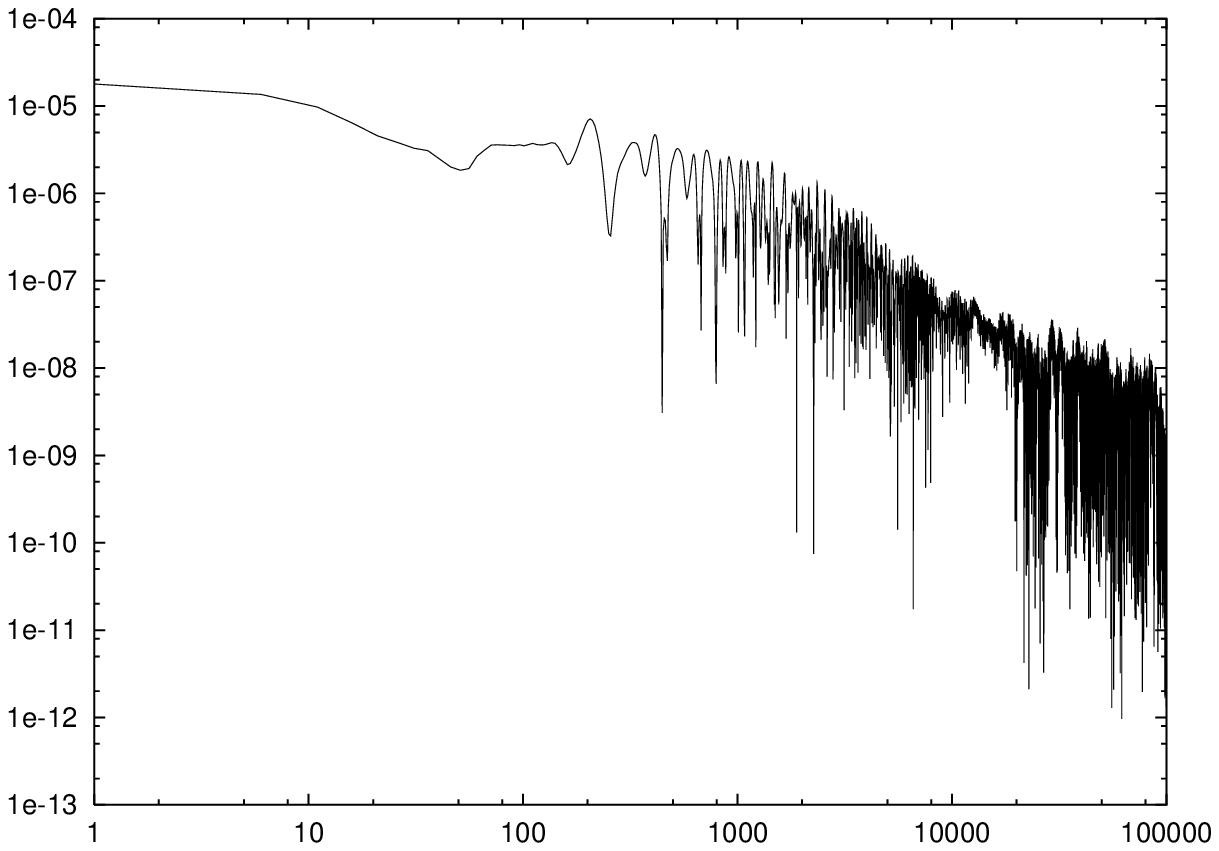}
  \includegraphics[width=0.4\textwidth]{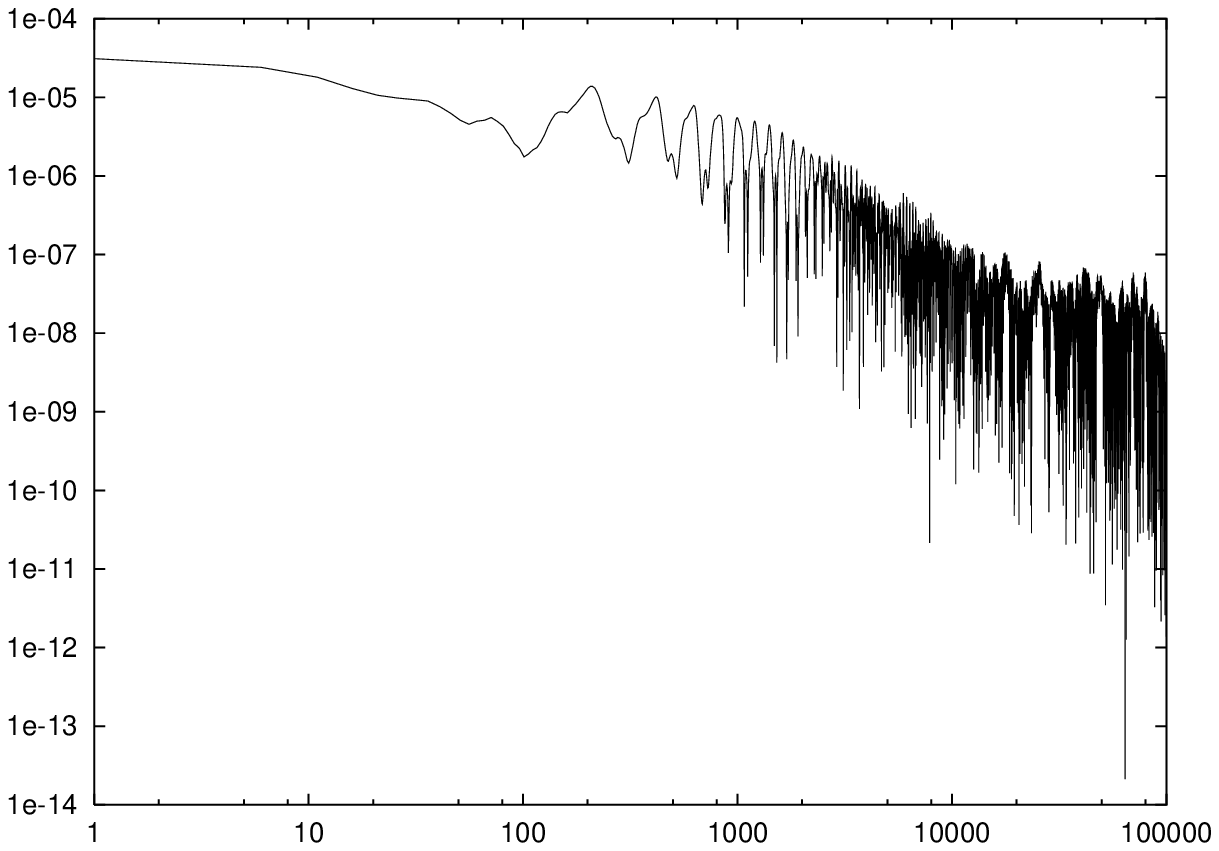}
  \includegraphics[width=0.4\textwidth]{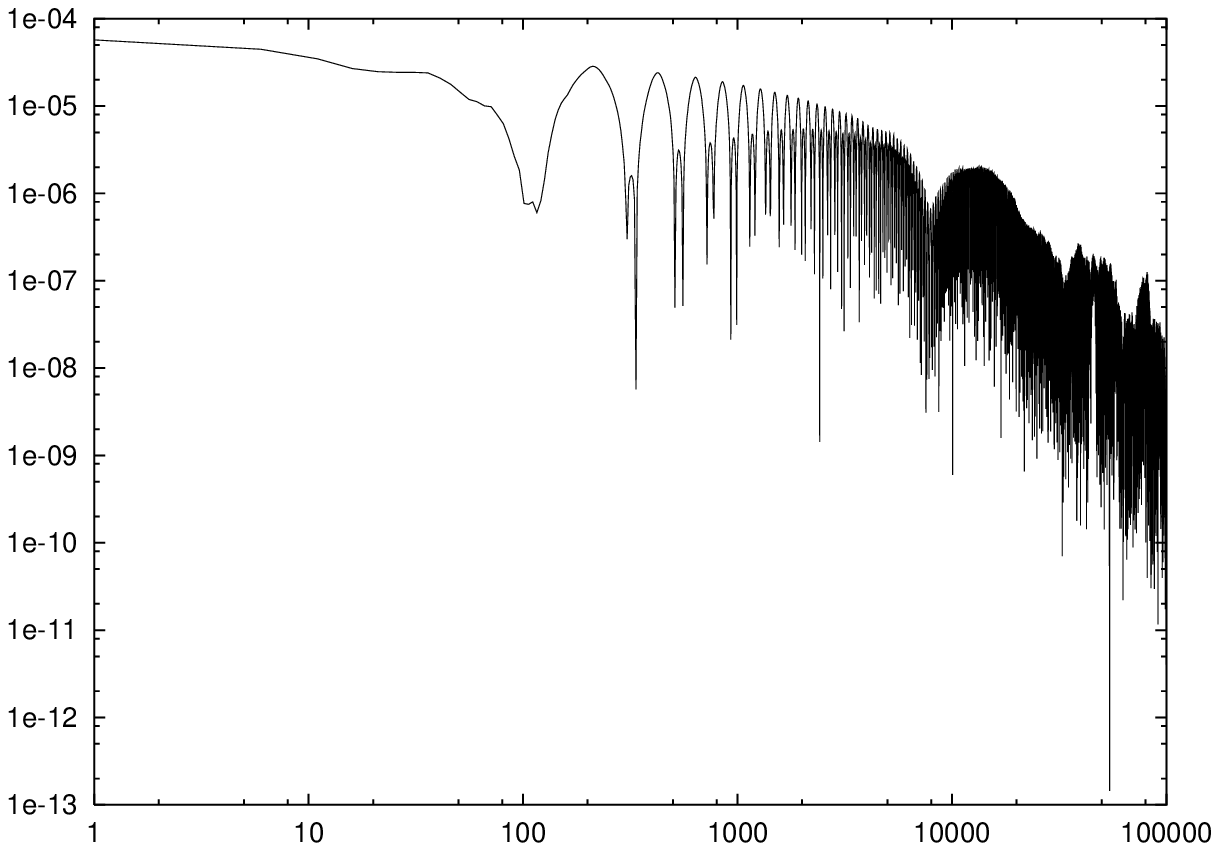}
  \caption{Heat flux autocorrelation function (log-log) of CNT $(10,10)$ 
    with 50, 100, 200 and 400 layers, respectively (from left to
    right, top to bottom).}
  \label{fig:log-log-hfacf}
\end{figure}

\fig{fig:log-log-hfacf} is the log--log version of \fig{fig:hfacf}.
Please note that data points in the range of 10000---100000 are in the
order of $10^7 \sim 10^8$ and almost random errors. The origin of
these errors are mainly due to inaccurate velocity trajectories and
roundoff errors in floating operations. Data points in the range of
1---1000 correspond to initial decay in very early moment and the
number of points is small (though this section seems quite long in the
graph), thus the middle section in the range of 1000---10000 time step
is most significant. and it can be seen that roughly the correlation
function decays as power law $f(t)=ct^\alpha$.

\begin{figure}[htbp]
  \centering \includegraphics[width=0.45\textwidth]{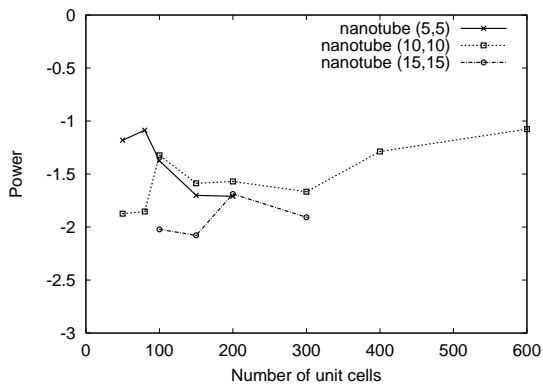}
  \caption{Relationship between decay power index and length of CNT
    $(5,5)$, $(10,10)$ and $(15,15)$. (Power indices for shorter
    CNT have also smaller errors.)}
  \label{fig:power-length}
\end{figure}

From data in \fig{fig:log-log-hfacf} we calculate the power index
$\alpha$ of heat flux autocorrelation function decay using linear
interpolation method and showed the relationship between $\alpha$ and
the length of CNT in \fig{fig:power-length}. By using the
mode-coupling theory, the knowledge of the asymptotic behavior of
$\langle \mathbf{J}(t)\mathbf{J}(0)$ allows determining the dependence
of thermal conductance on the system size $N$ \cite{lepri}. From
\fig{fig:power-length} it can been seen that the power index of heat
flux autocorrelation function decay is about $-3/2$, and for certain
cases it is near to $-1$, thus the thermal conductance $\kappa$ should
converge to a finite value as the system size increases. However, when
the length of CNTs is between $50 \sim 600$ layers in our work,
there is no evidence that $\kappa$ will converge.

\begin{figure}[htbp]
  \centering
  \includegraphics[width=0.45\textwidth]{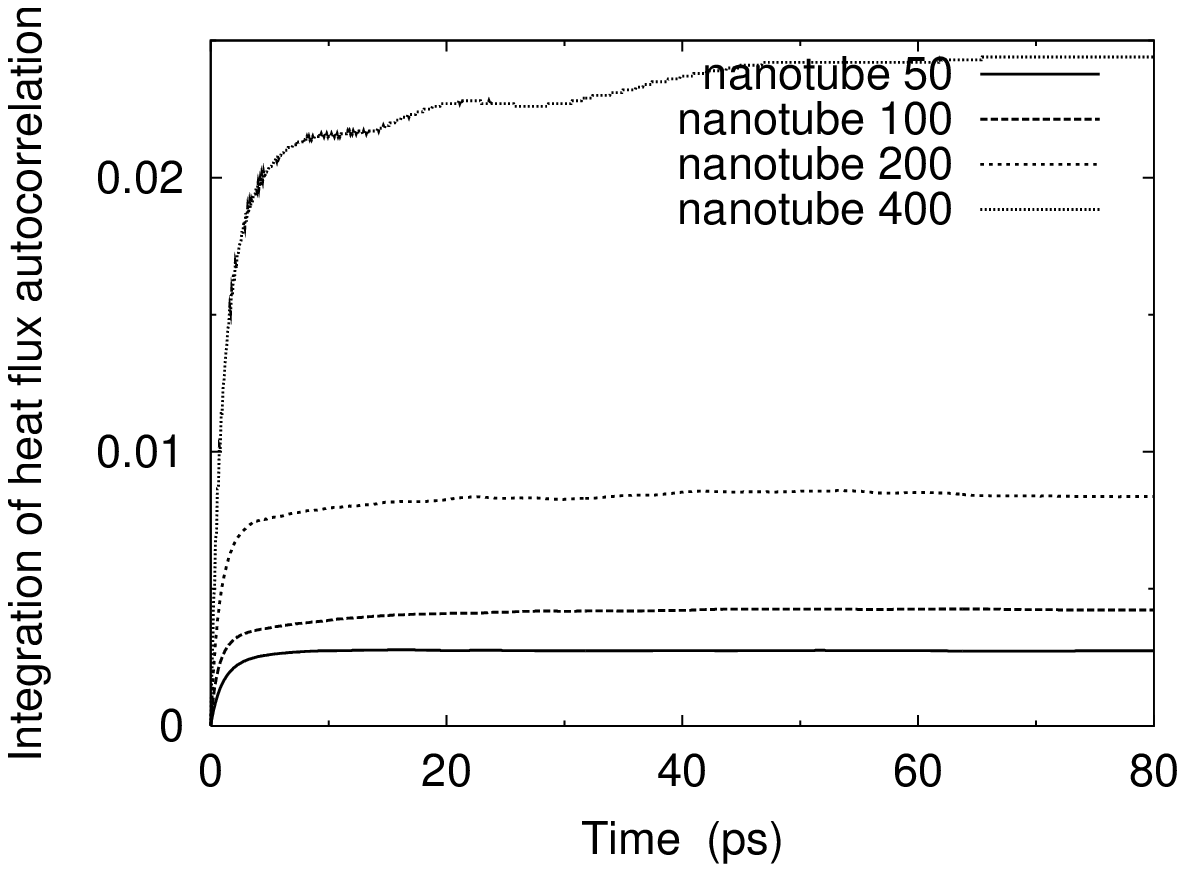}
  \includegraphics[width=0.45\textwidth]{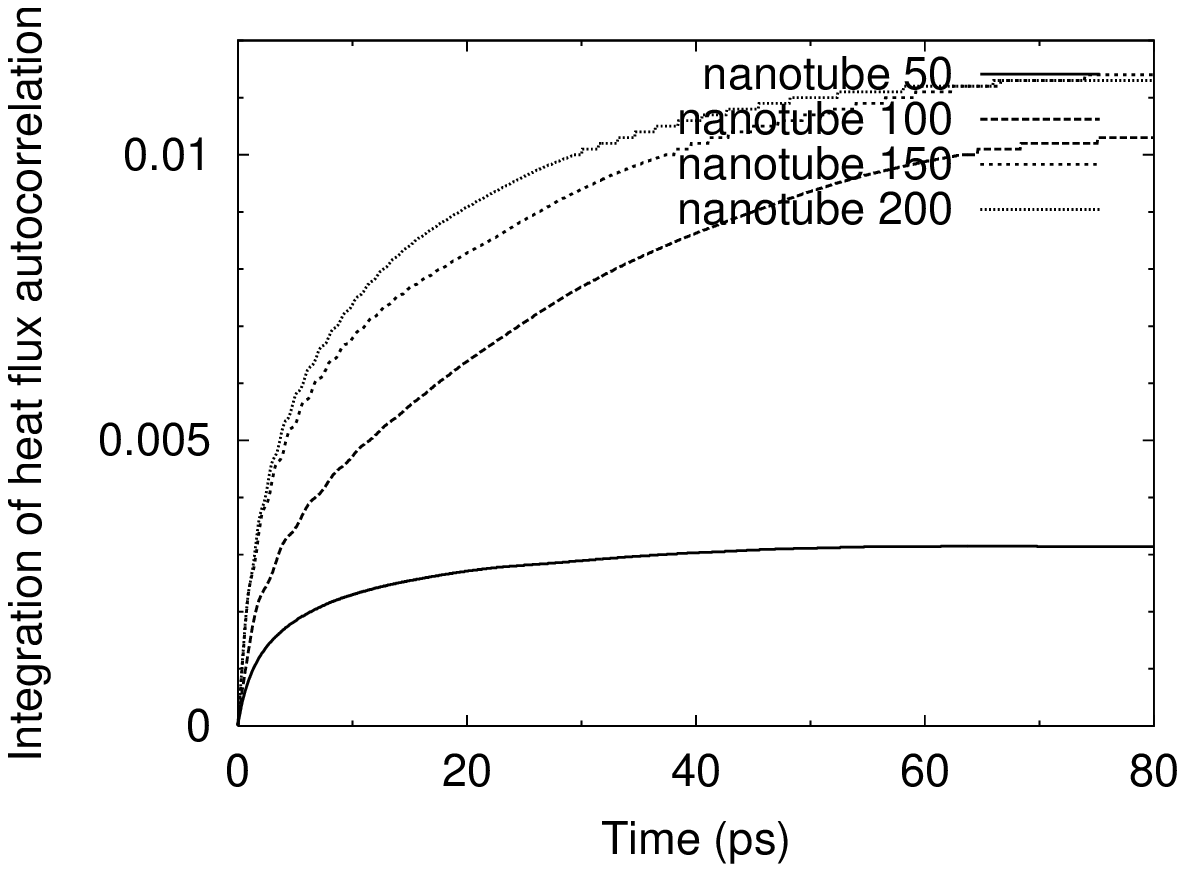}
  \caption{Integration of heat flux autocorrelation function over
    time $t$ up to given number of time steps for CNT $(10,10)$ and
    $(5,5)$.}
  \label{fig:integ}
\end{figure}

Curves shown in \fig{fig:integ} are the integration of heat flux
autocorrelation function over time $t$. It can be seen that very slow
decay doesn't contribute the thermal conductivity in a significant
way, and initial fast decay due to long wavelength low frequency
vibrational modes contributes mostly.  Therefore, the cutoff of long
wavelength vibrational modes will significantly influence the final
result of thermal conductivity. Compared with CNT $(10,10)$, the
integration of heat flux autocorrelation function for CNT $(5,5)$
converges hardly, especially for the longer CNTs. The integration of
heat flux autocorrelation function for CNT $(15,15)$ is qualitatively
the same as CNT $(10,10)$, and it converges fast, so we don't show
it in \fig{fig:integ}.

\begin{figure}[htbp]
  \centering
  \includegraphics[width=0.45\textwidth]{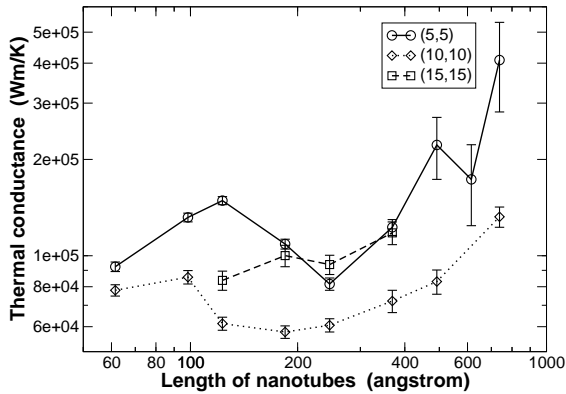}
  \caption{Thermal conductance as a function of CNT length. 
    In the graph solid line, dot line and dash line denotes the
    thermal conductance of CNT $(5,5)$, $(10,10)$, and $(15,15)$,
    respectively.}
  \label{fig:kappa-len}
\end{figure}

As the discussions in section \ref{sec:thermal conductance}, the
absolute value of an isolated single-wall CNT is ambiguous because the
cross section are is not well defined, so we discuss only thermal
conductance. The relationship between the thermal conductance and the
length of CNT is shown in \fig{fig:kappa-len}. In all cases, the
thermal conductance of single-wall CNT doesn't converge to a finite
value as the increase of CNT length. In the figure we also give the
standard errors of thermal conductance results and mark in the error
bars. The error is calculated as follows,
\begin{enumerate}
\item In micro-canonical ensemble simulation, $N_{\text{tot}}$ number
  of time steps is carried out, thus $N_{\text{tot}}$ heat flux data
  points is collected.
\item Divide these heat flux data to $N$ groups, each group has $M$
  heat flux data points, and calculate the heat flux autocorrelation
  function and integration for each group, respectively.
\item The final result is taken to be the average of $N$ groups of
  results in the second step. Meanwhile the standard error of the
  result is given by,
  \begin{equation*}
    S_i = \sqrt{\frac{1}{N}\sum_{j=1}^M (\xi_j-\bar{\xi})^2},
    \qquad i=1,\cdots, N.
  \end{equation*}
\end{enumerate}
 
From \fig{fig:kappa-len} it can been seen that as long as the length
of CNT increases, the thermal conductance increases correspondingly,
and this trend has been discovered in literature \cite{maruyama}. If
we consider the van der Waals thickness $3.4$ \AA as the thickness of
CNT shell, and treat CNT as a hollow cylinder, the thermal
conductivity results are in good agreement with S.~Maruyama's data
\cite{maruyama}.

\begin{figure}[htbp]
  \centering
  \includegraphics[width=0.45\textwidth]{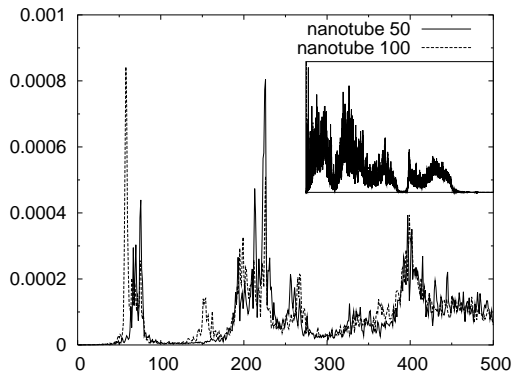}
  \caption{Vibrational density of states of CNT $(10,10)$ with
    50 and 100 layers, respectively. Inset shows the VDOS in the full
    frequency range, and the full graph shows the VDOS in the low
    frequency range. In the graphs, dash lines denotes VDOS of CNT with
    100 layers and solid line denotes VDOS of CNT with 50 layers.}
  \label{fig:dos}
\end{figure}

In order to know why longer CNT has higher thermal conductance, we
calculate the vibrational density of states (VDOS) by computing the
power spectrum of velocity correlation function while the simulation
is running, and the calculation can be expressed as follows
\cite{mutto},
\begin{equation}
  \label{eq:pdos}
  D_z(\omega)= \int\exp(-i\omega t) \langle v_z(t)v_z(0) \rangle dt,
\end{equation}
where $D_z(\omega)$ denotes the VDOS along the $z$ axis (i.e., the
axial direction), $v_z(t)$ denotes the velocity of atoms in the $z$
axis. \fig{fig:dos} shows the VDOS of CNTs with 50 and 100 layers,
respectively. The inset of \fig{fig:dos} shows the VDOS of two CNTs in
full frequency range, and it seems that two curves are almost the
same, this indicates that the middle and high frequency distribution
of VDOS of two CNTs are roughly identical. However, from the graph of
low frequency range, it can been seen that CNT with 100 layers has more
low frequency vibration modes, this is why longer CNT has higher
thermal conductance. We believe that why CNT with 100 layers has more
low frequency modes is that it has larger simulation domain and then
has longer phonon mean free path.

\section{Conclusions}
\label{sec:cons}
In this paper the high thermal conductance of single-wall CNTs is
calculated using equilibrium MD, and the relationship between thermal
conductance and length of CNT is discussed. It is found that as a kind
of quasi one dimensional material, CNT's thermal conductance doesn't
converge to a finite value as the CNT length increases up to 80 nm.
It can also be seen that longer CNT has more long wavelength
vibrational modes, and these modes contribute to the thermal
conduction as the CNT is longer.

\begin{acknowledgments}
  This project is supported by the Singapore-MIT Alliance. The authors
  thank Dr.~Gang Zhang for discussion and Dr.~Min Cheng for revising
  the manuscript.
\end{acknowledgments}


\end{document}